\documentstyle[11pt,aaspp4,flushrt]{article}
\tighten

\def\EROHR10{ERO~J164502+4626.4}
\def\sq{$\sqcup$\llap{$\sqcap$}}      

\def\spose#1{\hbox to 0pt{#1\hss}}
\def\iras{{\it IRAS }}

\def\lya{\ifmmode {\rm\,Ly\alpha}\else ${\rm\,Ly\alpha}$\fi}
\def\halpha{\ifmmode {\rm\,H\alpha}\else ${\rm\,H\alpha}$\fi}
\def\Mdot {\ifmmode {{\dot M}} \else ${{\dot M}}$\fi}

\def\kms{\ifmmode {\rm\,km\,s^{-1}}\else
    ${\rm\,km\,s^{-1}}$\fi}
\def\kmsMpc{\ifmmode {\rm\,km\,s^{-1}\,Mpc^{-1}}\else
    ${\rm\,km\,s^{-1}\,Mpc^{-1}}$\fi}

\def\msun{\ifmmode {\,M_\odot}\else ${\,M_\odot}$\fi}
\def\Msun{\ifmmode {\,M_\odot}\else ${\,M_\odot}$\fi}
\def\lsun{\ifmmode {\,L_\odot}\else ${\,L_\odot}$\fi}
\def\Lsun{\ifmmode {\,L_\odot}\else ${\,L_\odot}$\fi}
\def\rsun{\ifmmode {\,R_\odot}\else ${\,R_\odot}$\fi}
\def\Rsun{\ifmmode {\,R_\odot}\else ${\,R_\odot}$\fi}

\def\cm{{\rm\,cm}}
\def\cm3{\ifmmode {\rm\,cm^{-3}}\else ${\rm\,cm^{-3}}$\fi}

\def\ergps{\ifmmode {\rm\,erg\,s^{-1}}\else ${\rm\,erg\,s^{-1}}$\fi}
\def\ergpscm2{\ifmmode {\rm\,erg\,s^{-1}\,cm^{-2}}\else
    ${\rm\,erg\,s^{-1}\,cm^{-2}}$\fi}

\def\eg{{e.g.}}
\def\deg{\ifmmode {^{\circ}}\else {$^\circ$}\fi}
\def\degr{\ifmmode {^{\circ}}\else {$^\circ$}\fi}
\def\degs{\ifmmode {^{\circ}}\else {$^\circ$}\fi}

\def\etal{{et al.~}}

\def\h3Mpc{h^{-3}{\rm Mpc}^3}
\def\Ho{\ifmmode {\,H_0}\else ${\,H_0}$\fi}
\def\hnot{\ifmmode {\,H_0}\else ${\,H_0}$\fi}
\def\h0{\ifmmode {\,H_0}\else ${\,H_0}$\fi}
\def\hnotunit{\ifmmode {\rm\,km\,s^{-1}\,Mpc^{-1}}\else
    ${\rm\,km\,s^{-1}\,Mpc^{-1}}$\fi}
\def\qnot{\ifmmode {\,q_0}\else ${q_0}$\fi}
\def\q0{\ifmmode {\,q_0}\else ${q_0}$\fi}
\def\ie{{i.e.}}
\def\mic{\ifmmode {\rm\,\mu m}\else ${\rm \mu m}$\fi}
\def\micron{\ifmmode {\rm\,\mu m}\else ${\rm \mu m}$\fi}
\def\microns{\ifmmode {\rm\,\mu m}\else ${\rm \mu m}$\fi}

\def\arcsec{\ifmmode {^{\prime\prime}}\else $^{\prime\prime}$\fi}
\def\asec{\ifmmode {^{\prime\prime}}\else $^{\prime\prime}$\fi}
\def\arcmin{\ifmmode {^{\prime}}\else $^{\prime}$\fi}
\def\amin{\ifmmode {^{\prime}}\else $^{\prime}$\fi}

\def\secper{\ifmmode \rlap.{^{s}}\else $\rlap{.}{^{s}} $\fi}
\def\minper{\ifmmode \rlap.{^{m}}\else $\rlap{.}{^m} $\fi}
\def\magper{\ifmmode \rlap.{^{m}}\else $\rlap{.}{^m} $\fi}
\def\farcs{\ifmmode \rlap.{^{\prime\prime}}\else
    $\rlap.{^{\prime\prime}}$\fi}
\def\arcsper{\ifmmode \rlap.{^{\prime\prime}}\else
    $\rlap.{^{\prime\prime}}$\fi}
\def\arcmper{\ifmmode \rlap.{^{\prime}}\else
    $\rlap.{^{\prime}}$\fi}
\def\spose#1{\hbox to 0pt{#1\hss}}
\def\simlt{\mathrel{\spose{\lower 3pt\hbox{$\mathchar"218$}}
     \raise 2.0pt\hbox{$\mathchar"13C$}}}
\def\simgt{\mathrel{\spose{\lower 3pt\hbox{$\mathchar"218$}}
     \raise 2.0pt\hbox{$\mathchar"13E$}}}

\def\aa{{A\&A}}

\def\aasupp{{A\&AS}}
\def\aj{{AJ}}
\def\apj{{ApJ}}

\def\apjl{{ApJ}}
\def\apjsupp{{ApJS}}
\def\apjs{{ApJS}}

\def\mn{{MNRAS}}
\def\mnras{{MNRAS}}

\def\pasp{{PASP}}

\def\apjref#1;#2;#3;#4 {\par\pp#1, {#2}, #3, #4 \par}

\received{---}
\revised{---}
\accepted{---}
\articleid{}{}

\tighten

\lefthead{Dey et al.}

\begin{document}

\title{Observations of a $z=1.44$ Dusty, Ultraluminous Galaxy and Implications for Deep Sub-mm Surveys}

\author{Arjun Dey\altaffilmark{1,2}}
\affil{Department of Physics \& Astronomy, The Johns Hopkins University, 
Baltimore, MD 21218}
\affil{dey@noao.edu}

\author{James R.\ Graham}
\affil{Astronomy Department, University of California at Berkeley, CA 94720}
\affil{jrg@astro.berkeley.edu}

\author{Rob J.\ Ivison\altaffilmark{3}}
\affil{Institute for Astronomy, Dept.\ of Physics \& Astronomy, 
       University of Edinburgh, Blackford Hill, Edinburgh EH9 3HJ,
       Scotland, UK}
\affil{rji@roe.ac.uk}

\author{Ian Smail\altaffilmark{4}}
\affil{Dept. of Physics, University of Durham, South Road, Durham DH1 3LE, England, UK}
\affil{ian.smail@durham.ac.uk}

\author{Gillian S. Wright}
\affil{Institute for Astronomy, Dept.\ of Physics \& Astronomy,
       University of Edinburgh, Blackford Hill, Edinburgh EH9 3HJ,
       Scotland, UK}
\affil{G.Wright@roe.ac.uk}

\and

\author {Michael C.\ Liu}
\affil{Astronomy Department, University of California at Berkeley, CA 94720}
\affil{mliu@astro.berkeley.edu}

\smallskip

\centerline{\it Accepted for publication in the Astrophysical Journal}

\altaffiltext{1}{Hubble Fellow}
\altaffiltext{2}{Present address: NOAO, 950 N. Cherry Ave., Tucson, AZ 85719}
\altaffiltext{3}{PPARC Advanced Fellow}
\altaffiltext{4}{Royal Society University Research Fellow}

\vfill\eject

\begin{abstract}

We present new near-infrared and optical spectroscopic observations
which confirm the redshift of the $z=1.44$ extremely red object
ERO~J164502+4626.4 (object \# 10 of Hu \& Ridgway 1994; formerly known
as `HR~10' or `[HR94]~10') and a {\it Hubble Space Telescope} image
which reveals a reflected-S--shaped morphology at (rest--frame)
near-ultraviolet wavelengths.  The contrast between the rest--frame
far-red ($\lambda\lambda 8200-9800$\AA) and near-UV ($\lambda\lambda
2900-3900$\AA) morphologies suggests that the central regions of the
galaxy are heavily obscured by dust and that the galaxy is most likely
an interacting or disturbed system.  We also present new photometry of
this object at 450\micron, 850\micron\ and 1350\micron\ obtained using
the SCUBA submillimeter camera on the James Clerk Maxwell Telescope.
Our sub-mm data are extremely sensitive to emission from cold dust at
high redshift.  The rest--frame spectral energy distribution of
ERO~J164502+4626.4 is best understood in terms of a highly reddened
stellar population with ongoing star formation, as originally suggested
by Graham \& Dey (1996).  The new submillimeter data presented here
indicate that the remarkable similarity to ultraluminous infrared
galaxies (ULIRGs) such as Arp~220 and Mrk~231 extends into the
rest-frame far-infrared which bears the signature of thermal emission
from dust, presumably heated by young stars. ERO~J164502+4626.4 is
extremely luminous ($L\approx 7\times 10^{12}h_{50}^{-2}\,\lsun$) and
dusty ($M_{\rm dust}\approx 7 \times 10^8 (T_{\rm dust}/{\rm 40
K})^{-5} h_{50}^{-2} \Msun$). If its luminosity is powered by young hot
stars, then ERO~J164502+4626.4 is forming stars at the prodigious rate
of $\Mdot = 1000 - 2000~h_{50}^{-2} \Msun\, yr^{-1}$.  We conclude that
ERO~J164502+4626.4 is a distant analogue of the nearby ULIRG
population, the more distant or less luminous counterparts of which may
be missed by even the deepest existing optical surveys. The sub-mm
emitters recently discovered by deep SCUBA surveys may be galaxies
similar to ERO~J164502+4626.4 (but perhaps more distant). This
population of extremely dusty galaxies may also contribute
significantly to the cosmic sub-mm background emission.

\end{abstract}


\vfill\eject

\section{Introduction}

Near--infrared (near--IR) imaging surveys have resulted in the discovery of a
population of infrared--bright, extremely red objects (`EROs'), which
may be of significance to studies of galaxy evolution.  EROs, which we
define in this paper as having observed optical--near-IR colors
$R - K > 6$, have been identified both in the field, and around
high-redshift radio galaxies and quasars (Elston, Rieke, \& Rieke 1988,
1989, 1991; McCarthy, Persson, \& West 1992; Eisenhardt \& Dickinson
1992; Graham et al.\ 1994; Hu \& Ridgway 1994; Dey, Spinrad, \&
Dickinson 1995). Images of EROs from ground--based telescopes show that 
they are spatially extended on scales of $\sim
0\farcs5$. This suggests that EROs are probably galaxies rather than
stellar objects (e.g., brown dwarfs). Most EROs were not
{\em a priori} selected as radio sources and are generally radio quiet.
Extremely faint optically ($R > 24.5$), their distances and spectral
properties (i.e., whether they are normal or active galaxies) remain
unknown.

The extreme colors of these objects might be attributed either to an
old stellar population, or to a younger stellar population reddened by
dust (Graham \& Dey 1996, hereafter GD96; Yamada et al.\ 1997). The
spectral energy distributions (hereafter SEDs) of the reddest EROs
known are inconsistent with unreddened old populations at any redshift,
and suggest that they must be highly reddened starburst galaxies,
perhaps the distant counterparts of the local ultraluminous infrared
galaxies (ULIRGs; e.g., Soifer et al. 1984) discovered by the {\it
Infrared Astronomical Satellite} (\iras).

Most EROs have been discovered serendipitously, and their space
densities are therefore uncertain. Some measurements suggest that the
most extreme EROs (\ie, those as red as \EROHR10) are as abundant as
quasars, with a surface density in blank fields $\approx
0.01$\,arcmin$^{-2}$ (Hu \& Ridgway 1994; Cowie et al.\ 1994).  More
recent measurements suggest that the surface density of objects with
$R-K^\prime\ge 6$ and $K^\prime \le 17.5$ is $\approx
0.1\,$arcmin$^{-2}$ (Beckwith \etal 1998) with fainter EROs possibly
having even higher surface densities ($\approx 0.7\,$arcmin$^{-2}$ for
EROs with $K^\prime \le 20$; Eisenhardt \etal 1998).  There is also
some evidence that their surface density is higher in regions around
high-redshift radio galaxies and quasars compared with the general
field (e.g., Aragon-Salamanca et al.\ 1994; Dey, Spinrad, \& Dickinson
1995; Yamada et al.\ 1997).  An obvious interpretation of this result
is that the EROs are clustered around and hence physically associated
with the distant, luminous AGN, which would imply that EROs are at $z
\sim 1-3$ and their rest-frame optical luminosities are 5--20 times
more luminous than unevolved $L^*$ ellipticals at the same redshifts.
Alternatively, this excess number density may suggest that our samples
of distant luminous AGN are biased due to gravitational lensing by
foreground mass concentrations with which the EROs are associated.

Given that the EROs may constitute a significant population in our
Universe, it is important to understand their nature. In this paper we
present new optical, near-IR and sub-mm observations (\S~2) of
ERO~J164502+4626.4, an ERO discovered in the field of the $z=3.8$
quasar PC1643+4631 (object \# 10 of Hu \& Ridgway 1994)\footnote{This
ERO has been previously referred to as `HR~10' and `[HR94]' 10 (e.g.,
Graham \& Dey 1996; Cimatti \etal 1998). The nomenclature used in this
paper follows the IAU recommendations for source names.}.  Our new
observations convincingly demonstrate that \EROHR10\ is a $z=1.44$
luminous, dusty starburst galaxy, a distant counterpart of local ULIRGs
like Arp~220 and Mrk~231, as originally suggested by GD96 (\S~3). The
detection of \EROHR10\  at sub-mm wavelengths, and the relatively large
surface density of the ERO population suggests that these objects may
be candidates for the sub-mm emitters recently discovered in deep SCUBA
surveys (Smail \etal 1997, Blain \etal 1999a, Hughes \etal 1998, Barger
\etal 1998), and we briefly discuss this possibility and its
consequences in \S~3.5.

We assume $H_0=50h_{50}\,{\rm km\,s^{-1}\,Mpc^{-1}}$, $q_0 = 0.5$, and
$\Lambda=0$ throughout. For this cosmology, the luminosity distance of
\EROHR10\  is 10.53$h_{50}^{-1}$~Gpc. For \qnot=0.1, the luminosity distance
is larger by a factor of 1.286. 

\section{Observations and Results}

\subsection{HST WFPC2 Imaging}

The field of PC~1643+4631 was observed using the Wide Field Planetary
Camera 2 (WFPC2; Trauger \etal 1994, Holtzman \etal 1995) on the
refurbished {\it Hubble Space Telescope (HST)} through the F814W filter
on UT 1997 May 7. A position angle (P.A.) of 68.2\deg\ was used in
order to simultaneously observe both ERO~J164502+4626.4 and
ERO~J164457+4626.0 (objects \# 10 and 14 of Hu \& Ridgway 1994). Both
objects were targetted on the WF CCDs in order to optimize the
detection of faint diffuse emission. We obtained 4 exposures over 2
orbits with the telescope dithered by $\approx 0\farcs7$ between
orbits. The total integration time was 5300~sec.  The images resulting
from the calibration pipeline were corrected for cosmic rays,
registered and coadded. Flux calibration was performed using the
photometric zero points of Holtzman \etal (1995).  The final image
reaches a 3$\sigma$ limiting surface brightness of 26.77~AB~${\rm
mag\ arcsec^{-2}}$ in an 1\sq\arcsec\ aperture.

\begin{figure}
\plotfiddle{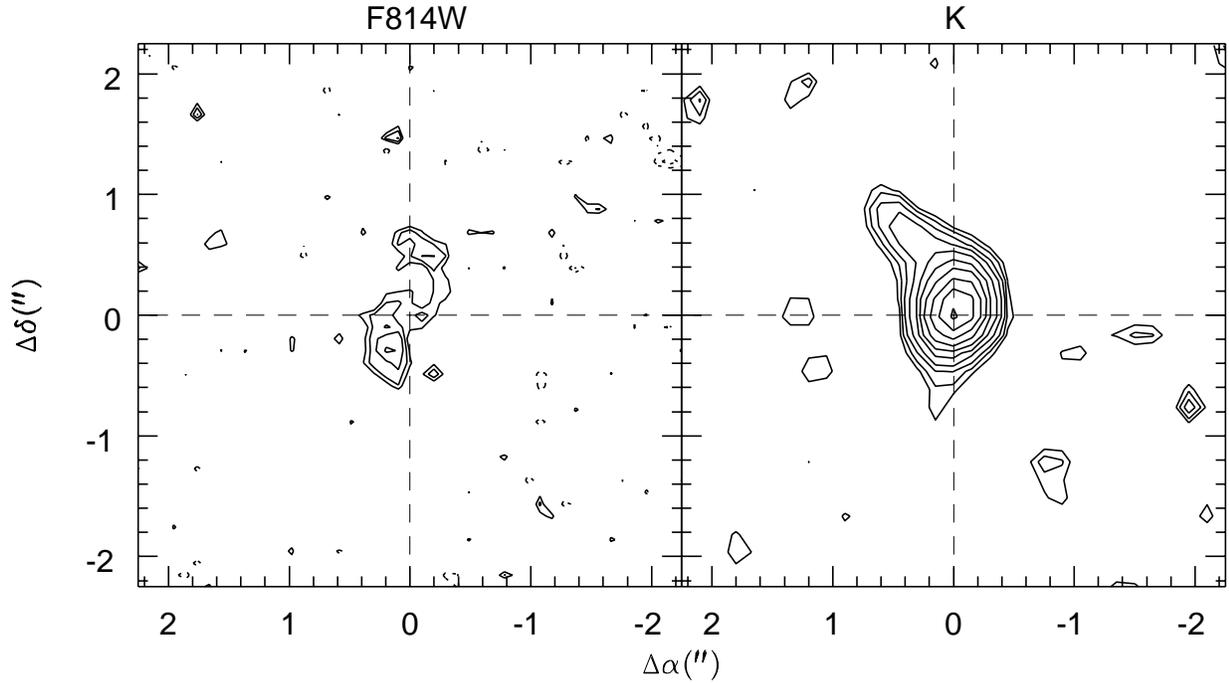}{5in}{0}{90}{90}{-270}{-170}
\caption{This figure presents a comparison of the observed $K$ and
F814W morphologies of ERO~J164502+4626.4.  The left panel shows a contour plot of
the {\it HST} WFPC2 image obtained through the F814W filter.  The
contour levels are drawn at levels (2,3,4,5,6,7,8,9,10)$\times\sigma_{sky}$, 
where $\sigma_{sky} = 25.56~AB\, {\rm mag\ arcsec^{-2}}$. Negative
contours are represented by dotted lines.  The right panel shows a
contour plot of the Lucy-deconvolved $K$--band image from Graham \& Dey
(1996).  The contour levels are drawn as in the left panel, but with
$\sigma_{sky}$ equivalent to $21.38\ {\rm mag\ arcsec^{-2}}$.  The
resolution of the F814W image is $\approx 0\farcs1$ whereas the
resolution of the Lucy--deconvolved $K$--band image is $\approx
0{\farcs}28$. Note that the morphologies are different and that the
peak of the $K$--band emission occurs in a region of low optical
emission.
\label{hr10cont}} 
\end{figure}

The WFPC2 F814W image of \EROHR10\ (rotated to a normal orientation
where north is up and east is to the left) is shown in
figure~\ref{hr10f814}. The galaxy is extended by roughly 0\farcs9 in
${\rm P.A.}\approx -21\deg$ in a reflected `S--shaped' morphology that
is suggestive of an interacting system or spiral / tidal arms. We
spatially coregistered the {\it HST} image with the near--IR $K$--band
image of GD96 using 16 common objects; the relative positional accuracy
is better than 0\farcs02. A comparison of the F814W and $K$--band
images reveals that the bulk of the near--IR emission arises from a
region of low optical flux, in between the two bright lobes of optical
emission (the ends of the reflected `S'; figure~\ref{hr10cont}). The
brightest region of optical emission is {\it not} cospatial with the
peak of the near--IR emission, but instead lies $\approx 0\farcs4$
south of it (figure~\ref{hr10cont}).

In a 3\arcsec\ diameter aperture, \EROHR10\  is found to have a
magnitude of $24.6 \pm 0.1$~AB mag (at $\lambda_{\rm obs}\approx
7930$\AA) or, equivalently, a Cousins $I$-band magnitude of $I_C\approx
24.2$. The color of \EROHR10\ in this aperture is then $I_C-K\approx
5.8\pm 0.1$, consistent with the values quoted by GD96 and HR94. 
The magnitudes of the northern and southern optical lobes,
measured in apertures of diameter 0\farcs5, are $26.4\pm 0.1$~AB mag
and $26.0\pm0.1$~AB mag respectively.

\subsection{Optical Spectrosopy}

We obtained optical spectra of \EROHR10\  on U.T.~1998 May 4 using the
Low Resolution Imaging Spectrometer (LRIS; Oke \etal 1995) at the
Cassegrain focus of the Keck II Telescope. The observations were made
in good seeing (FWHM $\approx 0\farcs6 - 0\farcs8$)
through a 1\farcs0 wide slit oriented in P.A. =
66.2\deg. LRIS was configured with the 400 line~${\rm mm^{-1}}$
grating ($\lambda_{blaze}\approx 8500$\AA; $\Delta\lambda_{\rm
FWHM}\approx 10$\AA), and the spectra cover the wavelength range
$\lambda\lambda5910-9730$\AA. The total exposure time was 3 hours, and the
target was dithered along the slit after every 30~min sub-exposure in
order to allow for proper fringe removal and sky subtraction. Wavelength
calibration was performed using NeAr lamps obtained immediately following
the \EROHR10\  observations, and relative flux calibration was performed using
observations of the standard stars Wolf~1346 and Feige~34 (Massey \etal 
1988, Massey \& Gronwall 1990). Although the
observations were made under non-photometric conditions, the spectrum
was scaled to an absolute scale using published photometry (GD96, Hu \&
Ridgway 1996).

\begin{figure}
\plotfiddle{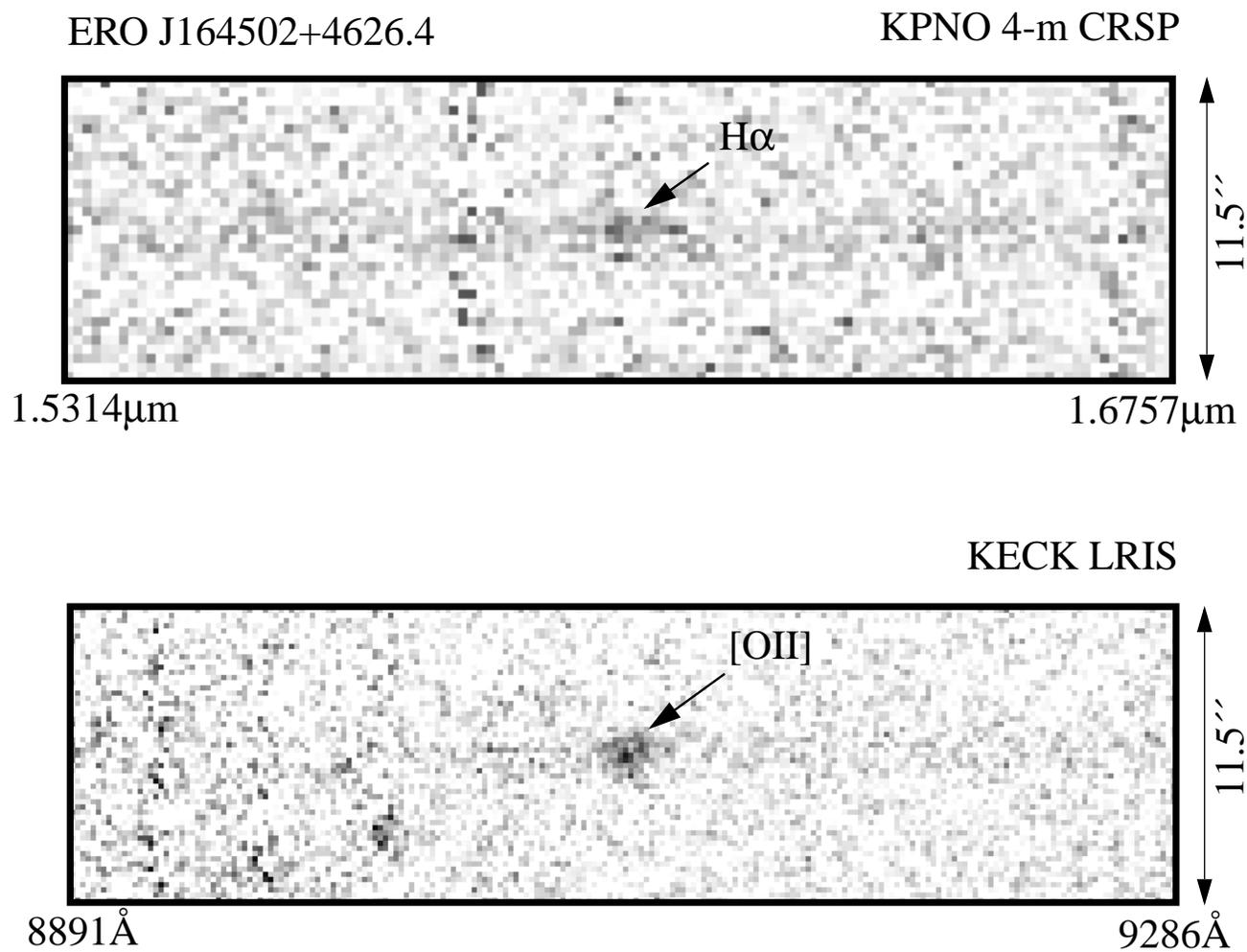}{5in}{0}{90}{90}{-270}{-170}
\caption{The above panels present portions of the two-dimensional
	spectra of ERO~J164502+4626.4 showing the emission line detections. The
	upper panel shows the detection of the H$\alpha$ emission line
	in the coadded near-IR spectrum obtained using CRSP on
	the KPNO 4-m telescope. The lower panel shows the detection of
	the [OII]$\lambda\lambda$3726,3729 emission doublet in the
	optical spectrum obtained using LRIS on the Keck II Telescope.
	\label{2dspectrum}} 
\end{figure}

\begin{figure}
\plotfiddle{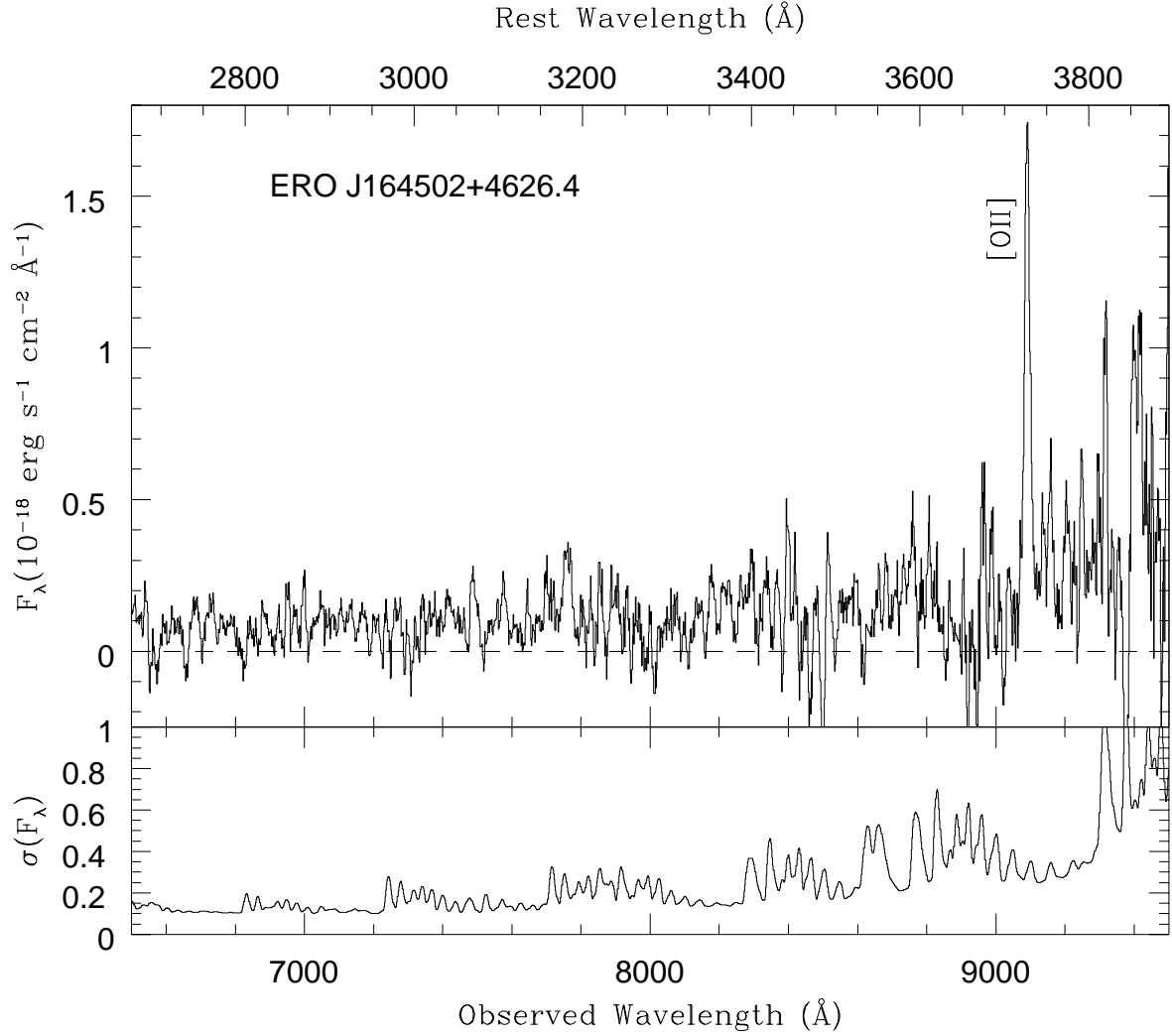}{5in}{0}{90}{90}{-280}{-170}
\caption{Optical spectrum of ERO~J164502+4626.4 obtained
	using LRIS on the Keck II Telescope. The upper panel shows
	the observed spectrum extracted in a $1\farcs0\times1\farcs7$
	aperture smoothed using a 13\AA\ width boxcar filter, and the 
	bottom panel shows the corresponding 1$\sigma$ error spectrum.
	The [OII]$\lambda\lambda$3726,3729 emission doublet is marked. 
	\label{optspectrum}}
\end{figure}

The sky-subtracted two-dimensional spectrum is shown in
figure~\ref{2dspectrum}, and the extracted one-dimensional spectrum
(extracted in a 1\farcs7 aperture) and the corresponding 1$\sigma$
errors are presented in figure~\ref{optspectrum}.  The notable features
in the optical spectrum are a strong emission line at
$\lambda$9090\AA\ (which corresponds to the
[OII]$\lambda\lambda$3726,3729 doublet at $z=1.439$) and a very red
continuum emission ($F_\nu\propto \nu^{-5.5\pm0.5}$ in the wavelength
range $\lambda\lambda 6500-9000$\AA). The signal-to-noise ratio in the
continuum is too poor to convincingly detect any absorption lines or
spectral breaks. The measured flux, FWHM and rest frame equivalent
width of [OII]$\lambda\lambda$3726,3729 are presented in
Table~\ref{emlines}.

\subsection{Near-Infrared Spectroscopy}

\begin{figure}
\plotfiddle{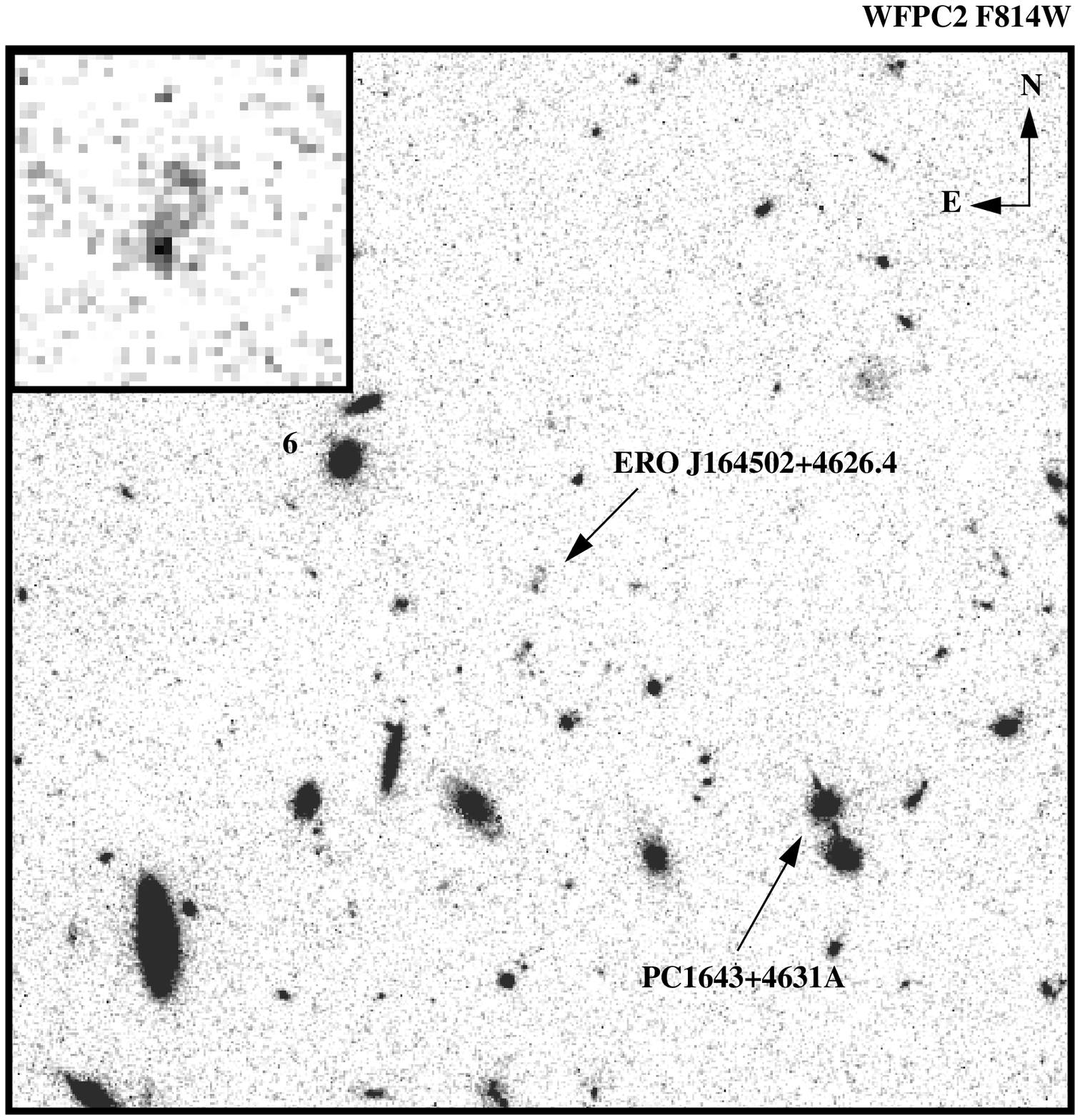}{6in}{0}{90}{90}{-280}{-100}
\caption{{\it HST} WFPC2 image of ERO~J164502+4626.4 (object 10 in 
	Hu \& Ridgway 1994) obtained through the F814W
	filter. The field of view shown is 50\arcsec\ on a side, and
	the ERO and the QSO PC1643+4631A are labelled.  North is up and
	east is to the left. ERO~J164502+4626.4 (``HR~10'')is located at
	$\alpha=16^h45^m02{\secper}36$, $\delta=+46\deg 26^\prime
	25{\farcs}5$ (J2000), and the offset from the QSO is
	$\Delta\alpha=+13{\farcs}85$, $\Delta\delta=+10{\farcs}27$. The 
        galaxy labelled ``6'' was used for aligning the dithered IR 
        spectra (see text). The
	inset, 4\arcsec\ on a side, shows ERO~J164502+4626.4.}
	\label{hr10f814} 
\end{figure}

Near-IR spectroscopy of \EROHR10\  was obtained using the Cryogenic
Spectrograph (CRSP; Joyce 1995) at the Cassegrain focus of the KPNO 4m
Mayall Telescope on the night of U.T.~1997 June 21. CRSP was configured
with grating \# 4 (200 ${\rm line\ mm^{-1}}$, $\lambda_{\rm
blaze}=3.0\micron$, dispersion ${\rm \approx 12\AA\ pixel^{-1}}$) used
in 2nd order and a 1\farcs0 slit to provide a resolution FWHM of 0.0042
\micron\ over the wavelength range 1.444\micron\ to 1.750\micron.
Observations of \EROHR10\  were obtained by dithering the object along
the slit; each dither sequence consisted of five equally spaced target
positions along the slit separated by 8\arcsec. The slit was oriented
at P.A.=59.3\deg\ in order to include a nearby galaxy (object 6 in
figure~\ref{hr10f814}; our notation follows that of HR94) to aid in
accurately coadding the data.  Each individual observation was 3~min
long, and the total exposure time was 195~min. The seeing during the
observations was typically 0\farcs8 in the $K$--band. The atmospheric
absorption was corrected using observations of the telluric standard
HR6064 (G1V) obtained at various airmasses both preceding and following
the \EROHR10\  observations.

The spectral frames were dark-subtracted, flat-fielded, rectified,
sky-subtracted, and coadded by using object 6 (see
figure~\ref{hr10f814}) to determine the accurate relative offsets. The
sky subtraction of each spectral frame was performed using a `local'
sky constructed from the median of the four preceding and four
following frames.  The summed two-dimensional spectrum is shown in the
upper panel of figure~\ref{2dspectrum}.  The extracted spectrum was
divided by the mean spectrum of the telluric standards to correct for
the telluric absorption, and corrected for the stellar features by
multiplying by the mean spectrum of a G0V star constructed by averaging
the spectra of HD109358 and HR5868 from the near-IR spectral atlas of
Lancon and Rocca-Volmerange (1992). Since the continuum emission from
\EROHR10\ is only marginally detected in these observations, the
telluric corrections are largely unnecessary, but also provide us with
a rough relative flux calibration. The resulting spectrum was then
scaled by a constant to be consistent with the $H$-band flux of
\EROHR10.

\begin{figure}
\plotfiddle{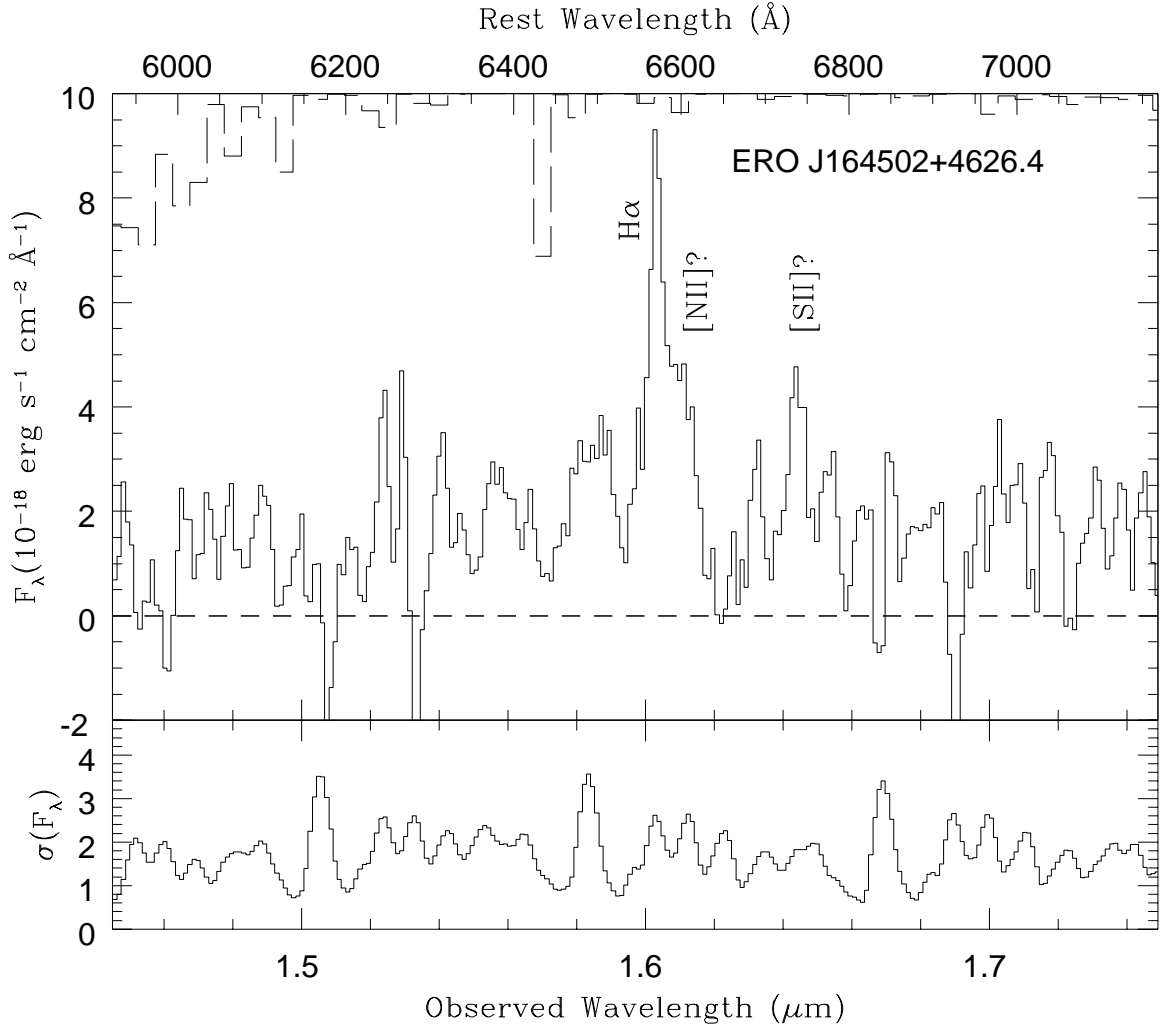}{5in}{0}{90}{90}{-280}{-170}
\caption{Near-IR spectrum of ERO~J164502+4626.4 obtained
	using CRSP on the KPNO 4-m telescope. The upper panel shows the
	observed spectrum extracted in a $1\farcs0\times1\farcs7$
	aperture and smoothed using a 36\AA\ width boxcar filter (solid line), 
	along with the relative atmospheric transmission (long dashed line). 
	The bottom panel shows the corresponding 1$\sigma$ error spectrum 
	which is dominated by the OH telluric emission lines. 
	The H$\alpha$ emission line and locations of the [NII]$\lambda$6584 and 
	[SII]$\lambda\lambda$6717,6731 emission doublet are marked.
	\label{irspectrum}}
\end{figure}

The calibrated near--IR spectrum of \EROHR10\ is shown in
figure~\ref{irspectrum}, and clearly shows the emission line detected
by GD96. The line is narrow, and appears to be resolved
into at least two components. The ratio of the wavelengths of the two
brighter components roughly matches the ratio of
[\ion{N}{2}]$\lambda$6584/\halpha, and is consistent with the original
identification (by GD96) of the line as {\halpha}+[\ion{N}{2}].  The extracted
spectrum also shows a weak feature at the expected location of the
[SII]$\lambda\lambda$6717,6731 doublet. However, given the
signal-to-noise ratio of the present data, we caution that the
[NII]$\lambda$6584 and the [SII]$\lambda\lambda$6717,6731 detections
are marginal, and should be treated accordingly (\eg,
figure~\ref{irspectrum}).

The emission line measurements are presented in Table~\ref{emlines}.
The numbers in this table were derived by using the SPECFIT software
(Kriss 1994) as implemented in {\it IRAF} to fit Gaussians to the
emission lines in the calibrated unsmoothed spectrum.  The H$\alpha$
and [NII]$\lambda$6584 emission lines were fit jointly by requiring the
two features to have equal width and holding the ratio of their central
wavelengths constant. The new high-resolution near-IR observation
presented here provides a more accurate measurement of the equivalent
widths and FWHMs of the H$\alpha$ and [NII] emission lines. In these
new higher spectral resolution data, the emission lines are found to be
narrower and have smaller equivalent widths than those derived by
GD96.

\subsection{Sub-mm Photometry}

Data were obtained during 1997 June, July and December, and 1998 January
and February using the Sub-mm Common-User Bolometer Array (SCUBA;
Holland et al.\ 1998). SCUBA has two arrays of bolometric detectors
which are operated at 0.1~K to achieve sky background-limited
performance on the telescope at 350--450 and 750--850\micron. Three
extra bolometers --- `photometric pixels' optimized for use at 1100,
1350 and 2000\micron\ --- are positioned around the long-wave array.

Photometry of sources significantly smaller than the beam are generally
performed using the central pixels of each array, which are aligned to
within an arcsecond of each other, or the photometric pixels. The best
photometric accuracy is achieved by averaging the source signal over a
slightly larger area than the beam, so the secondary mirror was `jiggled'
in a filled-square, 9-point pattern, covering $4\arcsec\ \times 4\arcsec$.
During the `jiggle' the secondary mirror was chopped azimuthally by
60\arcsec\ at 6.944\,Hz. After the first 9~sec `jiggle', the telescope was
nodded azimuthally to the reference position (subsequently every
18~sec). At 450\micron\ and 850\micron, we spent 260\,min on source, with a
further 165\,min at 1350\micron.

Skydips were performed before, during and after the target
measurements to determine the zenith opacities, and the telescope
pointing accuracy was checked regularly using 1633+382 and
3C\,345. All data were calibrated against observations of Mars and Uranus.
The observing conditions during the SCUBA runs were excellent, especially
during the early 1998 observing periods. 

Reduction of the 450\micron\ and 850\micron\ data consisted of taking the
measurements from the central bolometer, rejecting spikes, and averaging
over 18~sec time intervals. The signal detected by all the bolometers
is dominated by spatially correlated sky emission (Ivison et al.\ 1998a),
so data from the adjacent rings of bolometers were treated in a similar
manner.  The residual sky background was removed using the median of
the inner two rings of pixels, after rejecting those with excessive
noise. This reduced the noise-equivalent flux density at 850\micron\
to around 90\,mJy\,Hz$^{-1/2}$, suggesting that the effects of rapid
sky variability have been removed entirely.  Reduction of the 1350\micron\
data followed a similar process, though without the removal of sky noise.

At 850\micron, we find a clear detection of \EROHR10\ at a level well above
the expected extrapolation of the weak radio emission; at 450 and
1350\micron, the detection significance is lower, but the measured flux
densities are $> 3\sigma$ detections, and are entirely consistent with
thermal emission from optically thin dust. The flux density
measurements at 450\micron, 850\micron\ and 1350\micron\ are reported
in Table~\ref{submmdata}.

The 850\micron\ and 1350\micron\ flux densities reported in this paper
are roughly half the values reported by Cimatti \etal (1998).  Our more
extensive dataset was obtained during exceptional observing conditions,
as evident from the 450\micron\ detection, and we can exclude a
substantially higher flux with some confidence.  Furthermore, our data
were obtained using a single telescope with internally consistent flux
calibration, and the 450\micron\ and 850\micron\ photometric data were
obtained simultaneously. Hence, the relative photometry (and the
derived spectral index) are self consistent and more accurate than
other sub-mm measurements of \EROHR10\ presented to date.  There also
exist 95\micron\ and 175\micron\ observations of \EROHR10\ obtained
with the {\it Infrared Space Observatory} ({\it ISO}), which can
provide additional consistency checks on the sub-mm observations.  We
will report on these {\it ISO} observations in a forthcoming work by
Ivison \etal (1999).

\section{Discussion}

\subsection{Spectral Properties}

The spectroscopic observations presented in this paper confirm the $z=1.440$
redshift of the galaxy by resolving the emission line reported by
GD96 into two components which match the H-alpha + [NII]
identification, the detection of the
[OII]$\lambda\lambda$3726,3729 doublet, and the marginal detection of the 
[SII]$\lambda\lambda$6717,6731 emission lines at the same redshift.  

The H$\alpha$/[NII]$\lambda$6584 ratio ($\simgt 2.4$) and the
relatively narrow linewidths of the [OII] and H$\alpha$ emission lines
suggest that the ionization is more likely to be due to young, hot
stars than an active galactic nucleus (AGN). The spectrum of \EROHR10\ is
similar to that observed in local star--forming galaxies (\eg,
Kennicutt 1992). The FWHM  (deconvolved FWHM $\approx 260\kms$) of the
[OII] doublet and equivalent width of the H$\alpha$ emission line from
\EROHR10\ are more typical of star--forming galaxies than AGN (\eg, Liu \&
Kennicutt 1995).  However, since an AGN could be hidden behind dust and
rendered invisible at rest frame optical and ultraviolet wavelengths, a
firm statement on the nature of the energy source must await future
mid- and far-infrared spectroscopy. In the following discussion, we
assume that young stars power the entire far-infrared emission from
\EROHR10.

\subsection{Morphology}

The {\it HST\ } WFPC2 image (figures~\ref{hr10f814} and \ref{hr10cont})
clearly shows that \EROHR10\  
has an elongated and distorted morphology.  At a redshift of
1.44, the WFPC2 F814W filter samples the rest-frame wavelength range
$\lambda\lambda$2900$-$3900 (i.e., roughly rest-frame $U$-band light),
which includes the [\ion{O}{2}]$\lambda\lambda$3726,3729 emission
doublet. The equivalent width of this emission feature in \EROHR10\  implies
that the emission line contamination in the F814W filter is $\simlt$8\%;
the observed morphology is therefore dominated by the continuum emission
from the galaxy. Since the filter samples the continuum at near-UV
wavelengths just shortward of the 4000\AA\ break, the morphology
observed in the {\it HST\ } image only reflects the distribution of
UV  bright populations (\ie, hot, young stars), and in addition may be
heavily modified by the distribution of dust in the galaxy. The peculiar
UV morphology may be due to spiral arms, tidal features or something more
complex, and suggests that \EROHR10\  may be an interacting or distorted system. 

The most striking aspect of the rest frame near--UV morphology is that
it is very different from the rest frame far--red morphology observed
in the $K$--band: the red emission is more symmetric (at least at the
0\farcs6 resolution of the GD96 ground-based image) and peaks in a
region where the near--UV emission falls to a minimum, approximately at
the center of the reflected S--shaped structure seen in the WFPC2 image
(figure~\ref{hr10cont}).  Since the far--red emission in galaxies is
generally dominated by late type giant stars and an older main sequence
population and the near--UV emission is dominated by younger, hotter
stars, the observed morphological difference could be partly due to
spatially distinct populations, with the older stars being more
centrally concentrated (perhaps in a bulge?) compared to the younger
stars (perhaps in a disk or tidal arms?). 

Is it possible that the UV morphology is dominated by young super star
clusters?  The specific luminosities of the northern and southern lobes
observed in the near--UV ($\lambda_{rest}\approx 3250$\AA) image are
$5.4\times 10^{27}h_{50}^{-2}\,{\rm erg\,s^{-1}\,Hz^{-1}}$ and
$7.9\times 10^{27}h_{50}^{-2}\,{\rm erg\,s^{-1}\,Hz^{-1}}$
respectively. Therefore, these regions have luminosities that are at
least 3 times larger than the brightest of the super star clusters
observed in local starburst galaxies (\eg, Meurer \etal 1995), and it
is possible that these regions are composed of groups of star
clusters. We have not corrected the observed flux densities for the 
effects of extinction due to dust, and the luminosities quoted should be
regarded as lower limits.

However, since the extreme red color of \EROHR10\  requires a significant
amount of dust, the observed morphological difference could also be
produced by a spatially non-uniform dust distribution which reddens the
central regions more than the outer structure. Indeed, the colors in
any resolution element are redder than any reasonable stellar
population, implying that the entire visible spatial extent of the
galaxy is strongly reddened. It is intriguing that similar UV/optical
morphological differences are observed in the inner regions of some nearby
ULIRGs where they are commonly interpreted as resulting from spatial
differences in the dust extinction.

\subsection{Dust}

\begin{figure}
\plotfiddle{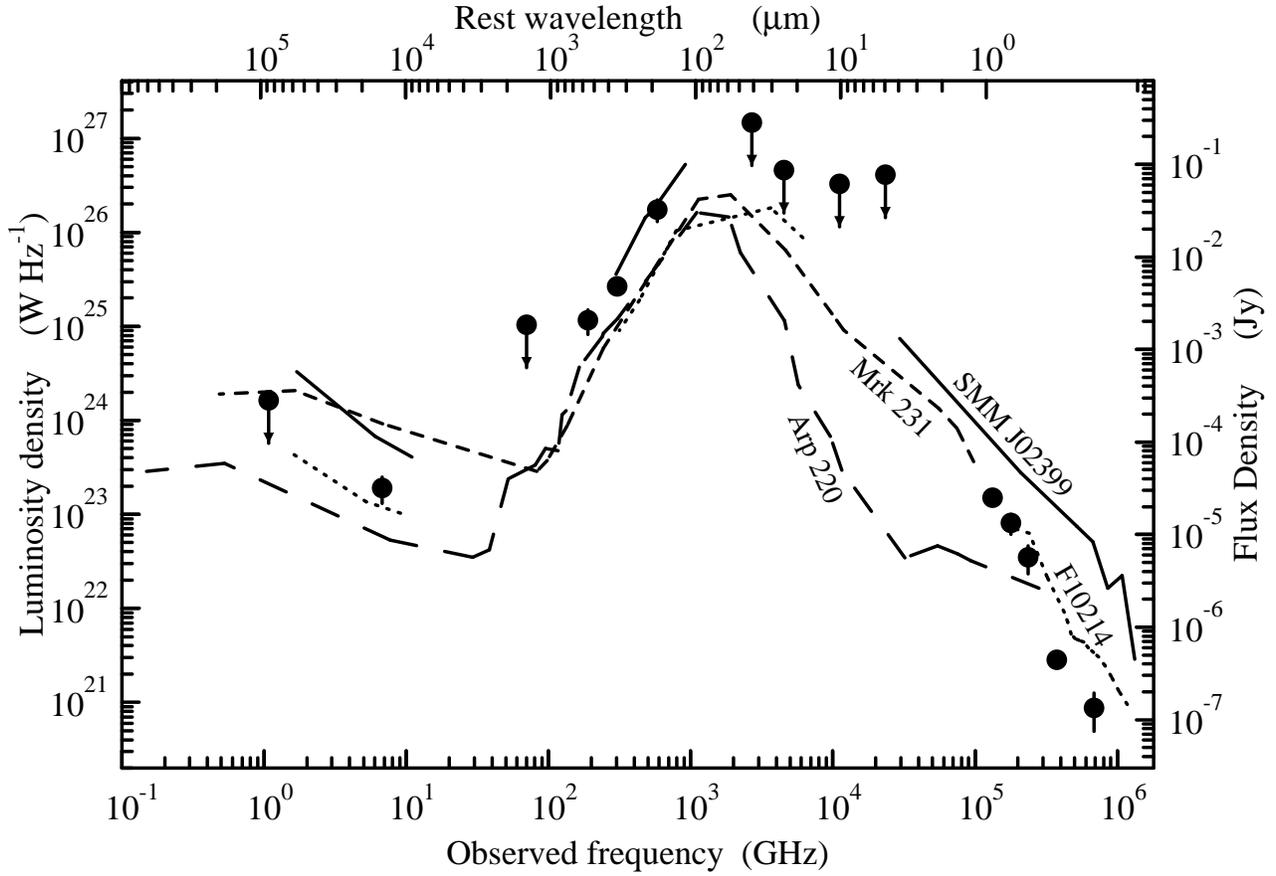}{4in}{0}{90}{90}{-280}{-150}
\caption{The SED of ERO~J164502+4626.4 between the radio and optical
	wavebands, represented by filled circles. The right-hand scale
	gives flux densities for ERO~J164502+4626.4. For comparison, we have plotted the SEDs
	of the ultraluminous {\em IRAS} sources F\,10214+4724
	(Rowan-Robinson et al.\ 1993; Barvainis et al. 1995), Mrk~231
	and Arp~220 (D.\,H.\ Hughes priv. comm.), and SMM~02399$-$0136
	(Ivison \etal 1998b) with units of luminosity density {\em
	(left-hand scale)}.  These lines are broken in regions where
	only upper limits are available.  For F\,10214+4724 and
	SMM~02399$-$0136 the SEDs are corrected for lensing by factors
	of 30 and 2.5 respectively (e.g., Graham \& Liu 1995;
	Broadhurst \& Leh\'ar 1996; Ivison \etal 1998b).}
\label{hr10sed}
\end{figure}

The SED of \EROHR10\  is remarkably similar to those of low redshift
ultraluminous galaxies like Arp~220 and Mrk~231 (figure~\ref{hr10sed}),
and this strongly suggests that \EROHR10\  is a dust--enshrouded system.  The
measured sub-millimeter flux densities at 450\micron, 850\micron\ and
1350\micron\ are greatly in excess of the power-law extrapolation of
the non-thermal radio emission, and it is therefore safe to assume that
the sub-millimeter flux results from thermal emission from dust heated
by an AGN or in star forming regions. The mm/sub-mm spectral index is 
also consistent with this hypothesis. 

\begin{figure}
\plotfiddle{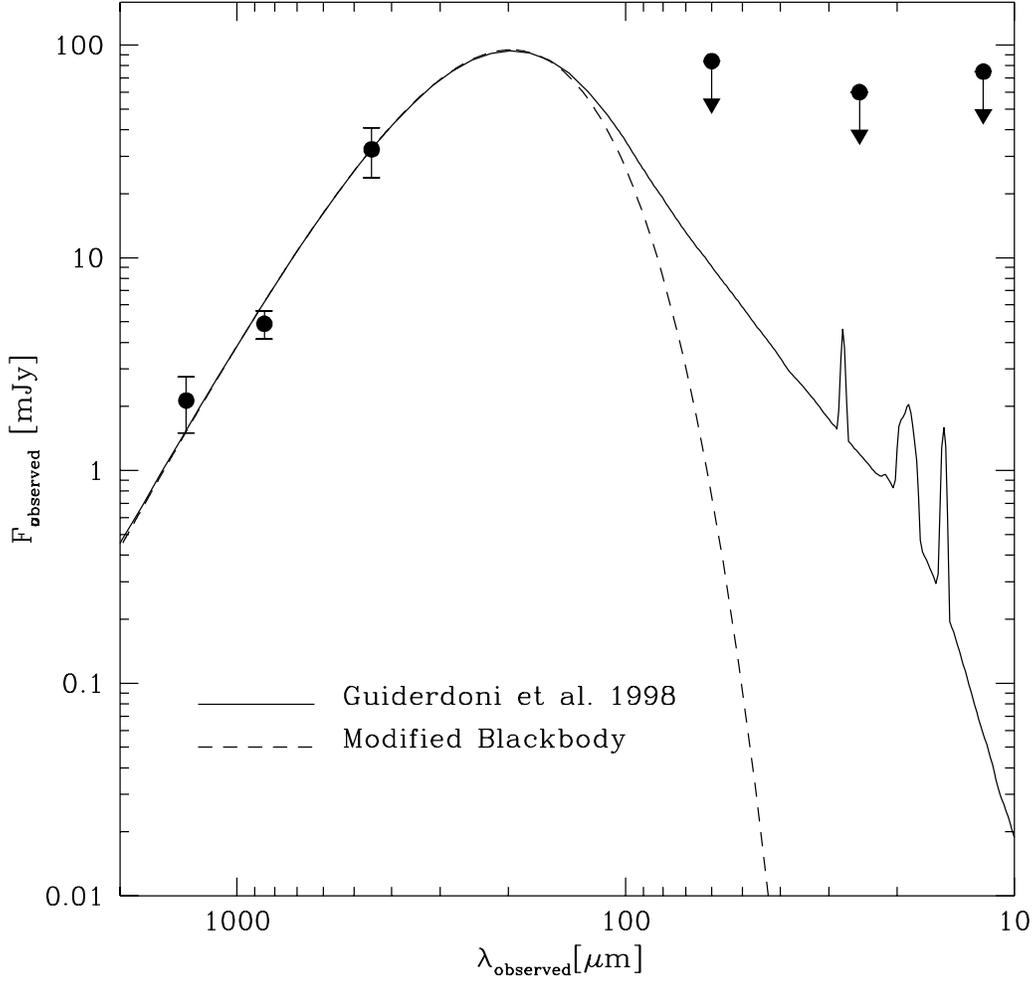}{4in}{0}{70}{70}{-210}{-100}
\caption{The rest frame far-infrared SED of ERO~J164502+4626.4 
	wavebands (filled circles) compared with a Guiderdoni \etal (1998)
	semi-empirical SED for IR luminous galaxies and a modified blackbody 
	($F_\nu\propto B_\nu[1-{\rm exp}\{-(\nu/\nu_0)^\beta\}]$) 
	of temperature $T=40$~K and emissivity index $\beta=1.5$. The models 
	shown were fitted to the three sub-mm flux density measurements.} 
\label{hr10sedmod}
\end{figure}

If we assume that the SED of \EROHR10\  is indeed identical to that of the
local ULIRGs, we can estimate the luminosity and mass of the warm dust
responsible for the thermal emission, and attempt to constrain the
star--formation rates necessary to power this galaxy.  Using the
template SED of an luminous {\it IRAS} galaxy compiled by Guiderdoni
\etal (1998), we find that a least squares fit of the Guiderdoni SED to
the three sub-mm points give a bolometric luminosity of $L = 7 \pm 1
\times 10^{12}\,h_{50}^{-2} \Lsun$ with reduced $\chi^2 = 1.1$
(figure~\ref{hr10sedmod}).  Almost all of this emission emerges in the
far infrared: $L_{\rm FIR}\approx 6.7\times 10^{12}\,h_{50}^{-2}
\Lsun$, where we have defined $L_{\rm FIR}$ as the luminosity emerging
between rest wavelengths of 10\micron\ and 2~cm.  This ranks \EROHR10\ 
among the most luminous (unlensed) IR galaxies known, and classifies it
as an ULIRG.

Note that the above estimate of the bolometric luminosity requires a
large extrapolation. The sub-mm luminosity (estimated from $\nu_{rest}
L_\nu^{rest}[850\micron/(1+z)]$, \ie, the best-determined sub-mm flux)
is $6\times 10^{10} \lsun$, and so this is an extrapolation of two
orders of magnitude. However, the sub-mm flux densities are well--fit
by optically thin thermal dust emission with an emissivity law with
index 1.5 (\ie, $Q_a \propto \lambda^{-1.5}$) and temperature of
$T_{\rm dust}=40 \pm 4$ K. This dust temperature is typical of ULIRGs
(Klaas \& Elsasser 1993). These modified black body model fits to the
sub-mm data have steeper spectra than the Guiderdoni \etal SED at
$\lambda_{\rm rest}\simlt 40~\micron$, and result in slightly lower
values of $L_{\rm FIR}$.  However, although the sub-mm data are all on
the Rayleigh-Jeans tail of the dust emisssion and therefore do not
strongly constrain the dust temperature, the small $\chi^2$ of the fit
and the reasonable value of the derived dust temperature suggest that
our estimate of the far-infrared luminosity is uncertain by at most a
factor of two.  The corresponding mass of dust is $M_{\rm dust}\approx
7 \times 10^8 (T_{\rm dust}/{\rm 40 K})^{-5} h_{50}^{-2} \Msun$ for
optically thin emission from grains with normal interstellar parameters
(Draine \& Lee 1984, Hildebrand 1983).

\subsection{Star Formation Rates}

The star formation rates can be estimated from either the
H$\alpha$ luminosity or the sub-mm continuum luminosity only if we assume
that young, hot stars provide the energy source for the line and
thermal continuum emission.

For our adopted cosmology (\hnot=50$h_{50}$\hnotunit\ and \qnot=0.5), the
luminosity in the H$\alpha$ emission line is $L_{\rm H\alpha} \approx
4.4\times 10^{42}\ h_{50}^{-2}\ {\rm erg\ s^{-1}}$. If this emission
line is powered entirely by young, hot stars, then its luminosity
implies a total star--formation rate of $\approx 40\ h_{50}^{-2}\ {\rm
\Msun\ yr^{-1}}$ (Kennicutt 1983). This is almost certainly a lower
limit to the true star formation rate in the galaxy, since we have not
made any corrections for the dust extinction.

A more reliable estimate for the star formation rate in \EROHR10\  can be
derived under the assumption that the sub-mm continuum excess is due to
thermal emission from dust grains heated by young, hot stars.  If stars
are formed at a constant rate over $10^7-10^8$~yr with a Salpeter
initial mass function ($\phi(m)\propto m^{-2.35}$; $0.1\le m \le
100$~\Msun), a luminosity of $10^{11} \Lsun$ corresponds to a star
formation rate \Mdot\ between ${\rm 14-24 \Msun\,yr^{-1}}$ (Leitherer
and Heckman 1995).  Therefore the observed bolometric luminosity
corresponds to a total star formation rate of $\Mdot = 1000 -
1800~h_{50}^{-2} \Msun\, yr^{-1}$ (or a factor of three smaller for the
formation rate of only massive stars).

A different estimate for the formation rate of massive stars can be
derived by relating the total luminosity to the mass consumption rate
in early type stars. Scoville \& Young (1983) derive $\Mdot_{\rm
OBA}\approx 7.7\times 10^{-11} (L/\Lsun) \Msun\, yr^{-1}$, which
implies a formation rate of $\approx 560 (T_{\rm dust}/40{\rm K})^5\,
\Msun\, yr^{-1}$ for the massive stars. Note that this estimate does
not account for the formation of lower mass stars, the mass cycled
through or locked up in lower mass stars and their remnants. This star
formation rate is therefore a lower limit, and as such is consistent
with the estimates derived above using the Leitherer \& Heckman (1983)
starburst models.

If the sub-mm luminosity is indeed powered by young stars, the derived
star formation rate is $26-45$ times greater than that inferred from
the H$\alpha$ luminosity; if this difference is due to dust extinction
by a foreground screen, it implies an extinction of $A_V\approx
4.5$~mag. For a Galactic extinction curve, this would imply that the
intrinsic ratio of [\ion{O}{2}] to H$\alpha$ of $(F_{\rm [OII]}/F_{\rm
H\alpha})_0 \approx 1.3$, which is at the extreme limit of observed
values for local star forming galaxies (\eg, Kennicutt 1992). Since the
dust is more likely to be intermixed with the line-emitting gas than
external to it, the extinction derived by comparing the sub-mm and
H$\alpha$ luminosities may not be directly applicable to the reddening
correction of the emission line ratios. The $A_V$
inferred here is about 2.5 mag larger than that estimated by fitting
the \EROHR10\  optical and near-IR flux densities with a reddened Sb
galaxy model SED (GD96).

The star formation rate derived from the sub-mm continuum emission is
extremely large and may lead us to the conclusion that the dust
emission is powered by an AGN rather than by star formation.  Indeed,
the source of the large luminosities in local ULIRGs is still a matter
of debate (\eg, Sanders \& Mirabel 1996).  However, there is no
evidence from the existing spectroscopy that this is the case: as noted
in \S~3.1, the emission line ratios and widths are more typical of
starburst galaxies than AGN.

The dust mass estimate from the previous section and the star formation
rate derived here can be combined to yield a crude order-of-magnitude
estimate for the lifetime of the starburst.  Assuming that \EROHR10\  has a
`normal' gas-to-dust ratio of 0.01, the total mass of gas is $\sim
7\times 10^{10}\Msun$, and the lifetime of the burst is $t_{\rm burst}
\simlt  1.3\times 10^8 $ yr. The total mass of stars produced is $\sim
10^{11}\,\Msun$, comparable to that of a present-day massive galaxy.

In this section we have assumed that the sub-mm continuum emission is
due to thermal emission from dust heated by young, hot stars. One
curious property of \EROHR10\  which may be relevant to this hypothesis is that
the rest frame 60\micron\ emission predicted by the dust emission and
the nonthermal radio emission at a rest wavelength of 6~cm
(extrapolated from the flux density observed at 1.4~GHz assuming a
$S_\nu\propto\nu^{-0.7}$ synchrotron spectrum) does not follow the
well-known 60\micron\ -- 6~cm correlation obeyed by local star forming
galaxies (\eg, de Jong \etal 1985). The rest frame 6~cm emission
predicted by this correlation for \EROHR10\  is $\approx 300\mu$Jy, more
than an order of magnitude above the detected radio emission
(Table~\ref{submmdata}).  The only ways of decreasing the predicted 
60\micron\ emission would be to decrease the dust temperature and increase 
the dust optical depth, both of which result in worse fits to the 
sub-mm data. The existing {\it ISO} data may be able to help constrain
the dust temperature and optical depth estimates (Ivison \etal 1999). 
This issue can also be resolved with better far-infrared
observations of {\EROHR10}, which will soon be possible with the 
{\it Space Infrared Facility} ({\it SIRTF}). 

If the large deviation from the 60\micron\ -- 6~cm correlation is real,
its interpretation is unclear.  It is unlikely that this deviation is
due to the sub-mm flux being heated by an AGN instead of a starburst,
since in the local universe AGN tend to depart from the correlation by
having excess nonthermal radio emission, not excess far-infrared
emission (\eg, Dey \& van Breugel 1994). Some local ULIRGs (\eg,
Arp~220) do show low frequency turnovers in their radio spectra which
are generally attributed to thermal absorption of the radio emission
(Sopp \& Alexander 1991). On the other hand, the first reliably
identified sub-mm selected high redshift galaxy, SMM~J02399$-$0136
($z=2.80$; Ivison \etal 1998b), has a `normal' 60\micron\ -- 6~cm ratio,
yet shows some evidence for an AGN contribution to the total emission.
Since the far-infrared flux densities are poorly constrained at present, 
and the origin of the correlation in local galaxies is also not
completely understood, we hesitate to draw any firm conclusions from
this observation.

\subsection{Cosmological Implications}

Measurements of the global star-formation history of the Universe,
using deep redshift surveys, e.g., the Canada France Redshift Survey
(Lilly et al. 1996) reaching $z \simeq 1$, the statistics of
Lyman-limit galaxies (Steidel \& Hamilton 1992) at $z=3.4$, the Hubble
Deep Field (HDF) $2.5 < z<4$ (Madau et al. 1996, Connolly et al. 1997), 
imply that the star-formation and metal-production rates were about 10
times greater at $z \simeq 1$ than in the local Universe, that they
reach a maximum somewhere in the redshift range $1\simlt z \simlt 1.5$ and 
remain roughly constant (or perhaps slowly decline) at higher redshifts
(Steidel \etal 1999). 

These conclusions, which are based almost entirely on samples selected at
optical and near-UV wavelengths, may be misleading (Smail, Ivison \&
Blain 1997; Blain \etal 1999a). Absorption by dust in regions of
star-formation may have distorted our picture of galaxy evolution in
the high- and low-redshift Universe in two ways.  First, neglect of
dust leads to an underestimate of the the star-formation rate in known
high- and low-redshift objects. Second, it is possible that an entire
population of heavily dust-enshrouded high-redshift objects, such as
EROs, has escaped undetected in the optical/UV surveys.

A useful diagnostic of the redness of the SED is the rest frame
infrared--to--blue luminosity ratio ($L_{FIR}/L_B$), where $L_{FIR} \equiv
\nu L_\nu$ at $\lambda_{\rm rest}=80\mu$m and $L_B \equiv
\nu L_\nu$ at $\lambda_{\rm rest}=4400$\AA. By this measure, the
reddest galaxy in the UGC is Arp~220 ($L_{FIR}/L_B\approx 60$; Soifer
\etal~1984). The mean $L_{FIR}/L_B$ value for the ultralumimous sample
is 25 (Sanders \etal 1987), and the reddest ultraluminous galaxy in the
{\it IRAS Bright Galaxy Catalog} is IRAS~12112+0305, which has
$L_{FIR}/L_B\approx 70$.  If we estimate $L_B$ for \EROHR10\ by
interpolating the observed $I$ and $J$ fluxes, we find that \EROHR10\ has
$f_B\approx 2.3~\mu$Jy. Hence, 
$L_B\approx \nu L_\nu(4400{\rm \AA})\approx 2.2\times
10^{10}\lsun$, and $L_{FIR}/L_B\approx 300$!  The name ERO is therefore
justified for this remarkable object.

The relevance of \EROHR10\ to cosmology and galaxy formation depends
critically on the space density of the ERO population, which is highly
uncertain at present.  Under the assumptions that the EROs form a
homogeneous population and the redshift and sub-mm luminosity of \EROHR10\
is typical of this class, GD96 estimated that the space density of
these objects is $\rho_{\rm ERO} \approx 2.1 \times 10^{-4}
h_{50}^{3}~Mpc^{-3}~mag^{-1}$. Hence, EROs may be more abundant than
the {\it IRAS} ultraluminous galaxies and quasars by two orders of
magnitude.  If the star formation rate for these galaxies is similar to
that observed in \EROHR10\ (\ie, $\Mdot > 580~\Msun\,yr^{-1}$), the star
formation rate at $z=1.4$ associated with the ERO population is $\simgt
0.1 h_{50} \Msun\,yr^{-1}\,Mpc^{-3}$. This sub-mm estimate of the star
formation density exceeds the rest-frame UV estimates of the
high--redshift star formation rate (from the CFRS and HDF surveys) by
more than a factor of six. It is noteworthy that this crude estimate is
roughly consistent with the predictions of Pei \& Fall (1995) (see also
Hughes \etal 1998).

Deep surveys at 850\micron\ of the sub-mm sky have resulted in the
detection of several faint sources ($\simgt 4$~mJy), about 20\% of
which appear to have no obvious optical counterpart brighter than
$I\sim 25$ (Smail \etal 1998). 
It is possible that these `missing' optical
identifications are galaxies similar to \EROHR10, perhaps even fainter at 
optical wavelengths, but with similarly strong 
thermal dust emission at sub-mm wavelengths.
Since the cumulative surface density of the sub-mm continuum emitters
with 850\micron\ flux densities $\ge 4$~mJy is $2.4\pm1.0\times
10^3\,{\rm deg^{-2}}$, the surface density of the optically faint
fraction is $\sim 480\,\,{\rm deg^{-2}}$ (estimates here are based on
the surface density determined by Blain \etal 1999b; however, see Barger
\etal 1998 for a lower estimate).  The surface density of EROs is
highly uncertain: Hu \& Ridgway (1994) estimated that the surface
density of objects with $I-K\ge6$ and $K\le 19$ is $\approx 36\,\,{\rm
deg^{-2}}$, whereas estimates of the surface density of less extreme
EROs (with $R-K\ge6$) discovered serendipitously tend to be larger by
at least an order of magnitude (\eg, Dey, Spinrad \& Dickinson 1995;
Knopp \& Chambers 1997; Beckwith \etal 1998). The roughly comparable
space densities of these two populations suggests that it is possible
that EROs comprise a significant fraction, perhaps all, of the
optically faint sub-mm emitters discovered in recent surveys. 
A large fraction of both local ULIRGs and 
distant SCUBA sources (Smail \etal 1998) appear to be interacting or 
merging systems, perhaps similar to \EROHR10.

Since the sub-mm sources thus far detected at flux densities $\simgt
0.5$~mJy arguably account for upto 100\% of the 850\micron\ cosmic
sub-mm background (Smail \etal 1997, Blain \etal 1999), the ERO
population may comprise a substantial component of the resolved
emission. The relevance of the properties of \EROHR10\ for our general
understanding of both the ERO and sub-mm populations will hinge on
whether it represents a particularly high luminosity or low redshift
member of this group.

The very red SED of \EROHR10\ also implies that it will be extremely
difficult to detect the most actively star-forming galaxies at higher
redshifts, since they may be completely enshrouded by dust and
undetectable at optical and near-IR wavelengths. For instance, if
\EROHR10\ were at $z=5$, it would have a $K\approx 25.8$ and $I>32$
(\hnot=50, \qnot=0.5), and therefore undetectable in even the deepest
existing ground- and space-based surveys. In contrast, the large
negative sub-mm k-correction would imply that the 850\micron\ flux
density of such an object would be $\sim 3$~mJy, comparable to that at
$z=1.44$.  The existence of dusty star-forming systems at $z\simgt 5$
is not an unrealistic expectation: the $z=5.34$ galaxy 0140+326RD1 (Dey
\etal 1998) has been recently argued to be a dusty star--forming system
based on its red $I-J$ color (Armus \etal 1998).  It is important to
note, however, that the inferred dust content of 0140+326RD1 is much
lower than that of \EROHR10: the implied extinction is $A_V\approx
0.5$~mag for 0140+326RD1, compared with $A_V\approx 4.5$ for \EROHR10.
Investigating the properties of the higher redshift counterparts of
systems like \EROHR10\ will require sensitive ground-based sub-mm
facilities with high-angular resolution and space-based
telescopes operating at near-IR and mid-IR wavelengths. In the near-future, 
the {\it Space Infrared Telescope Facility} ({\it SIRTF}) will provide 
flux density observations of EROs at mid- and far-infrared wavelengths  
($\sim 3-180$~\microns) allowing us to place more critical constraints
on the dust masses, temperatures and source luminosities, and perhaps 
determine the relationship of these objects to present-day galaxies. 

\section{Conclusion}

We have presented new optical, near--IR and sub-mm observations of
ERO~J164502+4626.4 (object \# 10 in Hu \& Ridgway 1994), 
an extremely red object in the field of the QSO PC~1643+4631A.
The new optical and near--IR spectroscopy show that \EROHR10\ is a distant
galaxy lying at a redshift of $z=1.44$. The peculiar rest frame
near--UV and far--red morphologies suggest that \EROHR10\ is a disturbed or
interacting system.  The far--IR and sub-mm evidence are consistent
with the hypothesis that this ERO is a dust-enshrouded object, with its
luminosity ($L\approx 7\times 10^{12}\,h_{50}^{-2}\,\Lsun$) powered by either a
starburst or an AGN. The existing spectral data for 
\EROHR10\ show no strong evidence for AGN emission at rest-frame near--UV or 
optical wavelengths. 
If the observed sub-mm continuum flux is due to
optically thin thermal emission from dust heated by a young, star
forming population, the inferred star formation rate is extremely large
($1000-2000\, \Msun\, yr^{-1}$).  In all its known properties, \EROHR10\
appears to be a high-redshift luminous counterpart of the dusty,
ultraluminous galaxies discovered in the local Universe by {\it IRAS}.
Although it is difficult to draw general conclusions from a single
object, less luminous or higher redshift galaxies similar to \EROHR10\ may
be missed from even the deepest existing optical and surveys.  It is
important to determine the space densities of these EROs and their
relevance to our understanding of galaxy formation and evolution in the 
distant universe.

\acknowledgements

We thank Brett Huggard, Gillian Rosenthal \& Dick Joyce for their
expert assistance during our KPNO observing runs, Richard Elston for
advice on using CRSP, the numerous people who have made possible the
process of obtaining and calibrating {\it HST} WFPC2 data, and the JCMT
staff for assistance during the SCUBA observations. We are grateful to
Wayne Wack, David Sprayberry and Bob Goodrich for their help with the
Keck LRIS observations. We thank Bill Reach for help with determining
the Galactic far-infrared foreground towards \EROHR10, 
and the referee,
Peter Eisenhardt, for constructive comments on our manuscript.  AD
acknowledges the support of NASA HF-01089.01-97A and partial support
from a Postdoctoral Research Fellowship at NOAO, operated by AURA, Inc.
under cooperative agreement with the NSF.  JRG acknowledges support
from NASA GO-06598.02-95A. The W.\ M.\ Keck Observatory is a scientific
partnership among the University of California, the California
Institute of Technology, and the National Aeronautics \& Space
Administration, and was made possible by the generous financial support
of the W.\ M.\ Keck Foundation.  The James Clerk Maxwell Telescope is
operated by The Joint Astronomy Centre on behalf of the Particle
Physics and Astronomy Research Council of the United Kingdom, the
Netherlands Organisation for Scientific Research and the National
Research Council of Canada.

\clearpage

\clearpage


 \begin{deluxetable}{lccccc}
 \footnotesize
 \tablewidth{0pt}
 \tablecaption{Emission Line Measurements of ERO~J164502+4626.4\tablenotemark{1}}
 \tablehead{ 	\colhead{Line} & 
		\colhead{$\lambda_{\rm obs}$} & 
		\colhead{Redshift} & 
		\colhead{Flux} &
		\colhead{FWHM} & 
		\colhead{$W_{\lambda,{\rm rest}}\tablenotemark{2}$}\nl
  & \colhead{\AA} & & \colhead{$10^{-17}{\rm erg\ s^{-1}\ cm^{-2}}$} & \colhead{\kms} & \colhead{\AA}}
 \startdata
 [OII]$\lambda\lambda$3726,3729 & 9090.6$\pm$0.9 & 1.439  & 2.6$\pm$0.4 & 420$\pm$100 & 47$\pm$5 \nl
 H$\alpha$      & 16030$\pm$7 & 1.443 & 33$\pm$8 & 597$\pm$140 & 89$\pm$20 \nl
 [NII]$\lambda$6584 & 16081.1\tablenotemark{3} & & 14$\pm$5 & 597\tablenotemark{3}   & 37$\pm$13 \nl
 [SII]$\lambda\lambda$6717,6731 & 16437$\pm$7 & 1.444 & 14$\pm$6 & 532$\pm$250 & 38$\pm$17 \nl
 \enddata
\tablenotetext{1}{All quoted measurements are based on Gaussian fits to the 
emission lines. The spectra were obtained through 1\farcs0 wide slits, and 
the spectral extractions used in these measurements are 
1\farcs7 wide in P.A.=66.2 ([OII]) and P.A.=59.3 (H$\alpha$, [NII] and [SII]).}
\tablenotetext{2}{Rest-frame equivalent widths assume $z=1.440$.}
\tablenotetext{3}{The central wavelength and width of the [NII]$\lambda$6584
emission line are fixed with respect to the derived values for the H$\alpha$
emission line.}
 \label{emlines}
 \end{deluxetable}

\begin{deluxetable}{ccccl}
\footnotesize
\tablewidth{0pt}
\tablecaption{Photometry of ERO~J164502+4626.4}
\tablehead{
\colhead{Observed} & \colhead{Rest} & \colhead{Flux Density} & 
	\colhead{Detector/} & \colhead{Reference} \nl
 \colhead{Wavelength} & \colhead{Wavelength} & \colhead{} & 
	\colhead{Instrument} & \nl
}
\startdata
  4400\AA   & 1800\AA & $ 0.16 \pm 0.07~\mu$Jy &  & GD96, Hu \& Ridgway (1992)\nl
  7930\AA   & 3250\AA & $ 0.52 \pm 0.06~\mu$Jy & WFPC2/{\it HST} & This paper \nl
 1.2\micron & 4920\AA & $ 6.4 \pm 2.1~\mu$Jy  &  & GD96, Hu \& Ridgway (1992)\nl
 1.6\micron & 6560\AA & $14.8 \pm 3.6~\mu$Jy  &  & GD96, Hu \& Ridgway (1992)\nl
 2.2\micron & 9010\AA & $27.7 \pm 0.6~\mu$Jy  &  & GD96 \nl
  12\micron & 4.9\micron & $<75$ mJy          & {\it IRAS}  & \nl
  25\micron & 10.2\micron & $<60$ mJy         & {\it IRAS}  & \nl
  60\micron & 24.6\micron & $<84$ mJy         & {\it IRAS}  & \nl
 100\micron & 41.0\micron & $<270$ mJy        & {\it IRAS}  & \nl
 450\micron &184\micron & $32.3  \pm 8.5 $ mJy  & SCUBA/JCMT & This paper \nl
 850\micron &348\micron & $ 4.89 \pm 0.74$ mJy  & SCUBA/JCMT & This paper \nl
1350\micron &553\micron & $ 2.13 \pm 0.63$ mJy  & SCUBA/JCMT & This paper \nl
  3.6~cm    & 1.5~cm  & $35 \pm 11~\mu$Jy  &      & Frayer (1996) \nl
  20~cm    & 8.6~cm  & $< 300~\mu$Jy &      & Frayer (1996) \nl
\enddata
\label{submmdata}
\end{deluxetable}

\end{document}